# Characterizing Transgender Health Issues in Twitter

**Amir Karami**
*University of South Carolina, USA. karami@sc.edu*

**Vanessa L. Kitzie**
*University of South Carolina, USA. kitzie@mailbox.sc.edu*

**Frank Webb**
*University of South Carolina, USA. fwebb@email.sc.edu*

## ABSTRACT
Although there are millions of transgender people in the world, a lack of information exists about their health issues. This issue has consequences for the medical field, which only has a nascent understanding of how to identify and meet this population's health-related needs. Social media sites like Twitter provide new opportunities for transgender people to overcome these barriers by sharing their personal health experiences. Our research employs a computational framework to collect tweets from self-identified transgender users, detect those that are health-related, and identify their information needs. This framework is significant because it provides a macro-scale perspective on an issue that lacks investigation at national or demographic levels. Our findings identified 54 distinct health-related topics that we grouped into 7 broader categories. Further, we found both linguistic and topical differences in the health-related information shared by transgender men (TM) as compared to transgender women (TW). These findings can help inform medical and policy-based strategies for health interventions within transgender communities. Also, our proposed approach can inform the development of computational strategies to identify the health-related information needs of other marginalized populations.

## INTRODUCTION
In recent medical research and practice, lesbian, gay, bisexual, transgender, and queer (LGBTQ) health has received increased attention. Compared to their heterosexual and cisgender peers, these populations face decreased rates of preventative health care, along with increased rates of HIV and depression (Gonzales, Przedworski, & Henning-Smith, 2016). These adverse outcomes are due, in part, to a dearth of information on LGBTQ health among both patients and medical professionals. For instance, half of the participants in a national survey of transgender identifying individuals reported a lack of provider knowledge about transgender care (James, Herman, Rankin, Keisling, Mottet, & Anafi, 2016). Existing research is limited by small samples that are difficult to use in generalizing results to a larger population. These small sampling frames reflect a lack of large-scale, national studies that focus on LGBTQ people; likely a result of these populations still being stigmatized and discriminated against at the federal level. Further, current survey methods that capture information about LGBTQ health do not collect data that can be compared longitudinally or to other data, the latter due to unstandardized collection methods. The survey method is also not efficient. For example, one survey on experiences of discrimination among transgender people took several years to develop and refine – from 2011 to 2016[1].

A unique challenge faced by transgender individuals is access to affirming health care, with 19% of those surveyed claiming they were denied care and 50% reporting the need to educate their provider on transgender health care needs (Grant, Mottet, & Tanis, 2011). This same report also indicates 41% of transgender people have attempted suicide and a have 2.64% rate of HIV infection, with higher rates for transgender women of color and those who are unemployed. A meta-regression study of existing surveys report that there are approximately one million transgender adults in the United States (Meerwijk & Sevelius, 2017); however, similar to LGBTQ health surveys at large, these surveys are not generalizable and tend to focus on specialized populations, such as college students and inmates. This lack of information for transgender individuals, and for all LGBTQ people, leads to significant limitations in adequately addressing the health disparities they face (Sell & Holliday, 2014).

Using social media, millions of individuals can share their everyday life and stories. Each day, approximately 500 million tweets are posted on Twitter discussing everyday life issues[2]. Social media facilitates people's belonging to and exchanging information within LGBTQ communities, by allowing users to transcend geographic barriers and contributing to the perception of limited risk in being "outed" (Byron, Rasmussen, Wright Toussaint, Lobo, Robinson, & Paradise, 2017).

---

[1] https://transequality.org/issues/national-transgender-discrimination-survey
[2] http://www.internetlivestats.com/twitter-statistics/

Privacy and stigma pose significant barriers to LGBTQ people sharing information related to these identities in clinical settings (Byron, Albury, & Evers, 2013). Therefore, social media data may provide unique perspectives on LGBTQ issues that are not shared in these other settings. Twitter is a popular social media platform from which to collect data because it offers Application Programming Interfaces (API) to collect large-scale datasets. Due to its APIs plus millions of users, several studies have used Twitter data to examine phenomena of interest in different domains such as health (Shaw & Karami, 2017; Webb, Karami, & Kitzie, 2018), psychology (Golder & Macy, 2011), library (Collins & Karami, 2018), business (Karami & Pendergraft, 2018), and disaster management (Cameron., Power, Robinson, & Yin, 2012).

This study is one of the few large-scale studies that examine the health-related information needs of transgender individuals on Twitter. To collect macro-level data, we developed a computational framework to obtain and analyze the tweets of self-identified transgender Twitter users to characterize the health issues they experience. In this way, we can begin to capture a broader picture of this population's health-related information needs. This study opens a new direction at the intersection of health informatics, social media analysis, and data science that provides practical benefits for medical and health experts. We organize the remainder of this paper as follows. In the next section, we review related work. Then, we present our framework and results in detail. Finally, we present a summary, limitations, and conclude with future directions.

**RELATED WORK**

The US Office of Disease Prevention and Health Promotion's Healthy People 2020 initiative identified LGBTQ people as facing significant health disparities including higher rates of substance abuse, mental illness and suicide attempts, and youth homelessness, as well as lower rates of health care utilization. Transgender individuals are the least likely to have health insurance among LGBTQ and heterosexual populations, and experience higher rates of suicide and victimization (Office of Disease Prevention and Health Promotion, 2017). These disparities are produced, in part, by the discriminatory treatment LGBTQ people face when receiving health care. Health care providers receive a cumulative median of five hours of LGBTQ-related curricular content during the whole of their medical instruction, and transgender individuals report being denied care due to their gender identities (Obedin-Maliver, Goldsmith, Stewart, and et al, 2011; Grant, Mottet, & Tanis, 2011). Together, evidence from these studies indicates a need for better educational resources for providers, as well as advocacy and protection for transgender rights to healthcare. Also, there are some unique health issues among transwomen and transmen such as breast cancer[3], sexual health (Reisner, Perkovich, & Mimiaga, 2010), pregnancy (Light, Obedin-Maliver, Sevelius, & Kerns, 2014); however, surveys have grouped transwomen and transmen into a larger "transgender" category, ignoring these important distinctions.

Healthy People 2020 identifies the lack of sexual orientation and gender identity information associated with medical records as an area of improvement to lessen disparities for LGBTQ populations. As of now, only two national data systems include information on gender identity (Office of Disease Prevention and Health Promotion, 2017). Efforts to collect the appropriate data focus on standardized terminology, confidentiality, and clarity of questions, as well as reducing barriers to patients' willingness to offer the information. Due to identification outside of a gender binary, even asking the question can be difficult due to multiple or no identities being applicable (Harrison, Grant, & Herman, 2012). By creating comprehensive records of gender identity associated with patients' medical records, including preferred pronouns, better education and treatment methods can be utilized to reduce disparities (Institute of Medicine (US) Board on the Health of Select Populations, 2013).

In addressing public health, the first step is assessing the situation by monitoring the health statuses and diagnosing the problems or hazards. Policy development follows this step (Harrell & Baker, 1994). By turning to social media, both the issues of scale and stigma for LGBTQ individuals can be addressed, allowing for large amounts of health data, as well as associated meta-data such as time and, in some cases, place, to be collected. Analysis of this data can inform the first step of addressing transgender public health issues by developing a framework of their information needs. This framework can be used to develop standardized large-scale data collection methods like surveys, as well as inform policy-based interventions.

Social media provides a new context in which to understand the unique experiences of sexual and gender minorities, including transgender people. Although prior research has used Twitter data for health applications, including intestinal disease (Zou, Lampos, Gorton, & Cox, 2016) and diabetes, diet, exercise, and obesity (Karami, Dahl, Turner-McGrievy, Kharrazi, & Shaw, 2018), these data have not been utilized to study the health information needs of transgender populations. This paper addresses this gap in the literature and proposes a computational framework using text mining methods to explore tweets of transgender users and identify the differences in the health-related information needs between TM and TW by analyzing their tweets using linguistics analysis and topic modeling.

---

[3] https://www.vumc.org/lgbti/key-transgender-health-concerns

## METHODLOGY AND RESULTS

We developed a framework to collect and analyze the content of the tweets by self-identified transgender users in Twitter. This framework has five steps: data collection, health data detection, linguistics analysis, and topic modeling (Figure 1).

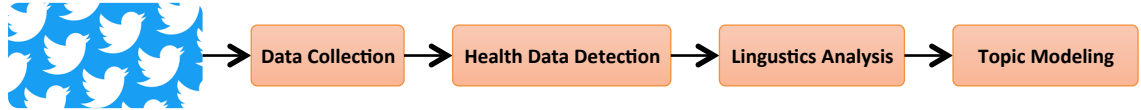

Figure 1: Research Framework

### *Data Collection*

To identify transgender users on Twitter, we searched profile descriptions for terms indicating identity disclosure (e.g., use of the label "FtM," which means "female to male"). Because transgender individuals comprise a vulnerable population, we avoid detailing all of our methods in this paper, however only focused on labels used explicitly to disclose a transgender identity. For this reason, we filtered the collected users to include those with at least 100 tweets and 100 followers, those identifying themselves as located in the United States, and those whose tweets were written primarily in English. This approach resulted in a dataset of 279 transgender users comprised of 73 self-identified TM and 206 self-identified TW users. We recognize that for many "transgender" represents a larger umbrella of non-cisgender identities (e.g., people identifying as "genderqueer"). For this paper, we wanted to focus on a subset of the population identifying as TM or TW since their needs might differ from other transgender-identified individuals who do not desire medical intervention. Focus on these other groups will constitute a future work. After we identified our 279 users, we used *twitteR* (a Twitter API within the statistical software package R) to collect up to 3,200 tweets for each of the 279 users. We removed retweets and the tweets containing URLs to focus content generated by these users, rather than the content they shared. This approach yielded 68,895 TM tweets and 215,556 TW tweets, for 284,451 tweets in total.

### *Health Data Collection*

From this dataset of 284,451 tweets, we then identified tweets containing health-related content. To make this detection, we applied the Linguistic Inquiry and Word Count (LIWC) software, which captures thoughts, feelings, personality, and motivations, to the dataset (Karami & Zhou, 2015; Karami & Zhou, 2014a). This tool has a health-related category showing whether a tweet contains health-related words. Using LIWC, we identified 5,492 TM tweets and 14,869 TW tweets for a total of 20,361 health-related tweets for analysis. Table 1 shows a sample of the collected health-related tweets. To maintain user privacy, we have lightly edited all tweets to avoid detection.

| | |
|---|---|
| 🐦 | *Recently received confirmation in writing from my health insurance that all transgender-related surgeries and services are now covered* |
| 🐦 | *Pretty much all day, I've been on a liquid diet* |
| 🐦 | *this sh\*tty neck pain I likely got from sleeping improperly is similar to my old neck injury* |
| 🐦 | *Eating, exercising, meds/ I've eaten, walked, done physical therapy, shopped, socialized, written, taken my meds and I've sewn!* |

Table 1. A Sample of Tweets

### *Linguistics Analysis*

After identifying health-related tweets from our 279 users, we then wanted to determine whether there were linguistic difference between health-related tweets shared by TM and TW. We used the five LIWC categories to make this determination. These categories are word count (i.e., number of words), analytics (i.e., degree of analytical thinking), clout (i.e., authority of author), authenticity (i.e., honest self-depiction), and tone (i.e., emotional inclination of author) (Arenas, 2018).

We applied two-tailed t-tests to compare TM and TW tweets based on the five categories. Table 2 shows that although there is no significant difference between TM and TW tweets with respect to tone, there are significant differences between the two groups with respect to the rest of categories. Findings indicate that TW users used more words in general, as well as more analytical, clout, and authentic-related words than TM users in their health-related tweets. This finding indicates that TW provide more details about their health issues than TM. One possible reason for providing more detail is that they have more trust in Twitter users than TM and want to obtain their feedback or suggestions.

|  | Word Count | Analytic | Clout | Authentic | Tone |
|---|---|---|---|---|---|
| P-value | 0.00<0.05 | 0.00<0.05 | 0.00<0.05 | 0.00<0.05 | 0.30>0.05 |

Table 2: P-value for Linguistics Analysis

*Topic Modeling*

When working with large datasets, it is difficult for people to manually identify health-related topics; there are simply too many tweets (Karami, 2017). To address this issue and understand what health-related topics transgender users were discussing, we applied Latent Dirichlet Allocation (LDA) to our dataset. This type of topic modeling is the most established and robust among the different types (Karami, Gangopadhyay, Zhou, & Kharrazi, 2015a). Researchers have used LDA in a wide range of applications such as analysis of health and medical corpora (Karami, Gangopadhyay, Zhou, & Kharrazi, 2018; Karami, Gangopadhyay, Zhou, & Karrazi, 2015b; Karami & Gangopadhyay, 2014), literature review (Priva & Austerweil, 2015), and politics (Karami, Bennett, & He, 2018; Najafabadi & Domanski, 2018; Najafabadi, 2017), as well as spam detection (Karami & Zhou, 2014b). For this research, we applied a Mallet implementation of LDA based on Java programming language (McCallum, 2002).

LDA assumes that each topic is a distribution over words and each document is a mixture of the topics in a corpus (Blei, Ng, & Jordan, 2002; Karami, 2015). The model categorizes the words that are semantically related in a topic to represent a theme. For example, this model assigns "gene," "dna," and "genetic" to a topic with the theme "Genetics" (Figure 2).

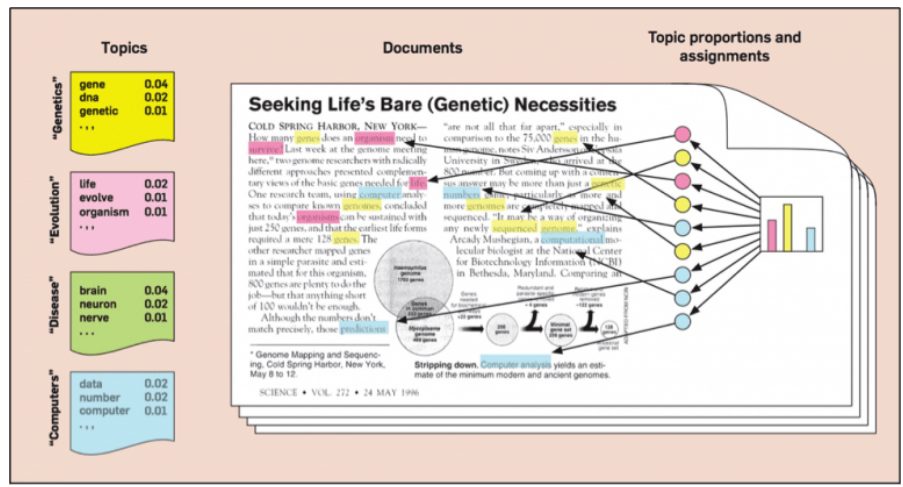

Figure 2: The Intuition Behind LDA (Blei, 2012)

Once we removed stopwords from our corpus (e.g., "and," "the"), we then determined how many topics best represented the data. To make this determination, we used the log-likelihood method to estimate the number of topics based on t inference of topic distribution on new, unseen documents and found the optimum number of topics to be 125 (Figure 3).

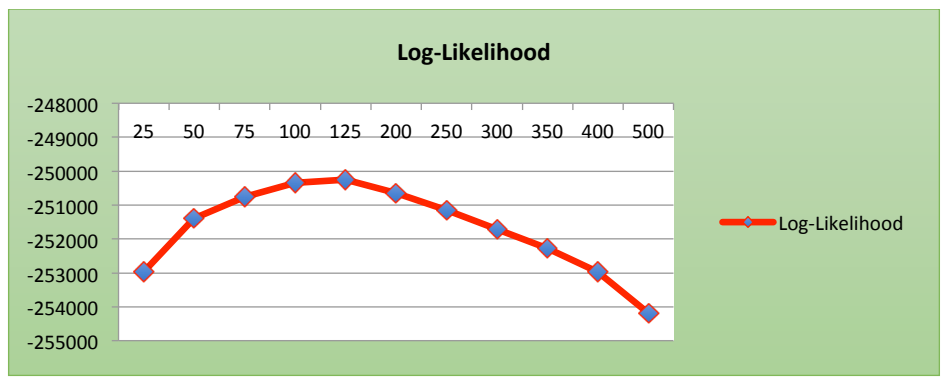

Figure 3: Determining Optimum Number of Topics

| T# | Topics | Label |
|---|---|---|
| T1 | surgery surgeon bottom chest ftm plastic feels grs recovery scars | Gender Confirmation (GC) Surgery |
| T2 | experience agree lives talking pain harm difficult personal feelings mind | Emotional State |
| T3 | exhausted today exhausting feel emotionally working physically home tiring bed | Emotional and Physical |
| T4 | live die work forever people disabled family act medicaid cuz | Health Legislation |
| T5 | mom home dad day hospital parents kids told family crying | Family's Health |
| T6 | great state blue pill states medicaid medicare parents social security | Health Legislation |
| T7 | pain dealing injury danger conversation patient coma addiction | Mental Health |
| T8 | tired people sick stressed scared ugh break angry wired boring | Mental Health |
| T9 | months hormones testosterone started ago hrt weeks change full transition | GC hormone replacement therapy (HRT) |
| T10 | care health insurance medical pay money cover plan cost bills | Health Legislation |
| T11 | life system death people support literally immune killing avoid operating | Critical Health Care |
| T12 | trans health safe care transgender rights support vital lgbtq hiv | General LGBTQ Health |
| T13 | free fat low class amazing creating culture planned milk water sugar | Diet |
| T14 | medical trans history research treatment professionals records doctors diagnosis condition | Clinical Health |
| T15 | type great lose alcoholic exercise arthritis weight injured surgeon diabetes | Diet and Lifestyle |
| T16 | hope great end time night headache pain stress lot infection | Symptoms |
| T17 | pretty exhausting awesome easy constantly medicine completely super benefit fairly | Positive Clinical Experience |
| T18 | doctor office appointment time called put message visit scheduled experience | Clinical Experience |
| T19 | calories burned walked miles workout exercise gym biking hill fitness | Exercise |
| T20 | estrogen dose testosterone hormone high shot low pills double level | HRT |
| T21 | work hard run exercise walk built makes fine time note | Exercise |
| T22 | sore feel hand pains muscle throat hurts shoulder back aches | Aches and Pains |
| T23 | life rest saved amazing entire transitioning celebrate changed worked peace | GC Positivity |
| T24 | terrible pill coughing shot aids band fit prob usual crisis flu | Illness and Symptoms |
| T25 | trans people transition hormones gender medical dysphoria surgery therapy doctors | GC Clinical |
| T26 | pain back cramps gain hip run suffering leg waking joint | Aches and Pains |
| T27 | therapist therapy work find doctor found thing clinic talk helped | Therapy |
| T28 | eyes face nose allergies running bleeding itchy summer painful chills blood | Allergies |
| T29 | gender identity diagnosis treat dysphoria specific informed ways hormones symptoms | GC Health |
| T30 | therapy good exercises session today bring therapist advocate trans working | Therapy |
| T31 | doctor side find testosterone prescription pharmacy effect estrogen bottle meds | GC Pharmaceutical |
| T32 | mental health illness people issues illnesses stigma chronic struggle month | Mental Health |
| T33 | care health listen speak abt taking doctors talking trust nice | Clinical Relation |
| T34 | cancer problem suicide kill die symptom breast depression suicidal attempt | Critical Health Care |
| T35 | fat body weight exercise loss size lazy mirror muscle face | Body Image and Gender Identity |
| T36 | hope healing power back pain helping wounds hurt fast prayers | Healing |
| T37 | blind hate turn eye operator broke glasses spot read touch | Vision |
| T38 | drink water coffee alcohol diet tea cigarette smoking coke caffeine | Lifestyle Habits |
| T39 | food eating diet poisoning sick addiction pizza chicken cheese chocolate | Food Poisoning |
| T40 | drugs drug thing alcohol weird fucked buying addict test illegal | Drug Use |
| T41 | physical emotional abuse hormones sexual violence labor practice drug hug | GC Stress |
| T42 | feeling sick feel sad nauseous crying depressed body fever dizzy | Symptoms and Stress |
| T43 | pregnant wife young family woman slowly age nurse ideas babies | Pregnancy |
| T44 | hormones doctor pills blood appointment clinic test excited waiting days | GC Positivity |
| T45 | things medical friend attack hospital apparently care bills panic heart | Health Finances Stress |
| T46 | pain depression anxiety medication chronic extreme disorder dysphoria bipolar ocd | Mental Health |
| T47 | brain stop science medicine term excuse injury vomit bigotry nonsense evidence | Variety of Symptoms and Society |
| T48 | part healing thing open reason normal works medication sad wound | Medical Reaction - Negative |
| T49 | bad hurt guy hit ache realize pain breaking felt stomach | Injuries |
| T50 | tired hours sleep night sick morning bad migraine wake feeling | Sleep |
| T51 | brain head headache migraine injuries bad migraines tumor deep breathing | Neurology - Headaches |
| T52 | surgery gender change therapist clinic sex hormones transition numbing surgeries | General GC |
| T53 | tired sleep night today tomorrow early woke insomnia stay bed | Sleep |
| T54 | painful honestly sounds allergic bad incredibly reaction allergy metal feeling | Allergies |

Table 3: Topics and Labels

From these 125 topics, 54 topics were relevant to health. Examples of irrelevant topics were those that discussed a health-related issue, but couched within a larger context, such as politics (e.g., "*trump, gop, abortion, president, debate*"). The middle column in Table 3 shows the top word for each of the topics that we found using LDA with 1000 iterations. Two members of the research team inductively labeled the topics based on the overall themes conveyed by the top words, then discussed and resolved any discrepancies in labeling (Table 3).

We compared these topics to other studies and surveys and found significant alignment. Specifically, the 2015 US Transgender Survey indicates that healthcare services such as insurance providers, transgender health care issues such as gender confirmation surgery and mental health are among the top concerns of transgender people (James, Herman, Rankin, Keisling, Mottet, & Anafi, 2016). Studies of transgender health issues also investigated diet, exercise, and disordered eating (VanKim, Erickson, Eisenberg, Lust, Simon Rosser, & Laska, 2014; Conron, Scott, Stowell, & Landers, 2012), sleep disorders (Shipherd, Green, & Abramovitz, 2010), drug use (Kecojevic, Wong, Schrager, Silva, Bloom, Iverson, & Lankenau, 2012), chronic pain disorders (Maurer, Lissounov, Knezevic, Candido, & Knezevic, 2016), body image (Atkins, 2012; Moradi, 2010), pregnancy (Light, Obedin-Maliver, Sevelius, & Kerns, 2014), and injury (House, Van Horn, Coppeans, & Stepleman, 2011).

Using the labels, we organized our 54 topics into seven categories including general healthcare, diet, healthcare services, mental health, symptoms, transgender healthcare, and wellbeing. Healthcare services and diet represent categories with the highest and the lowest number of topics, respectively (Figure 4):

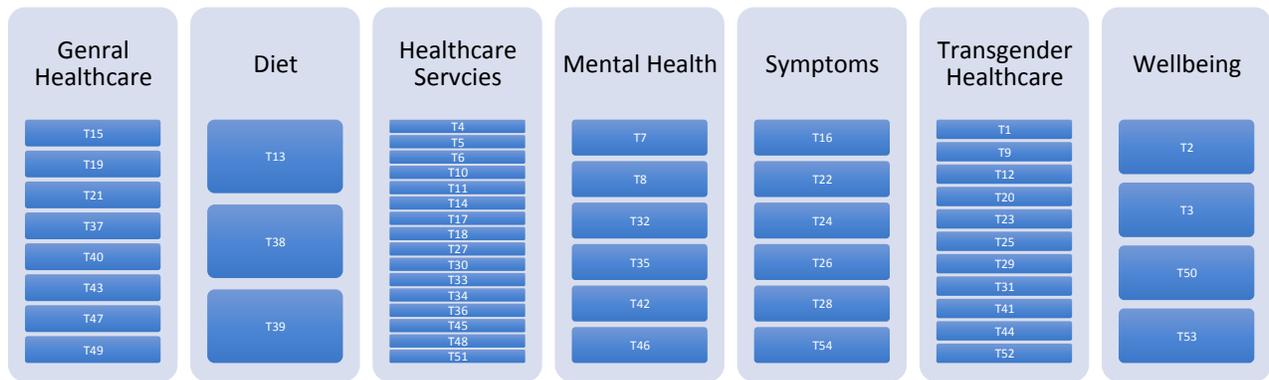

**Figure 4: Categories of Topics**

- **General Healthcare**: The category describes general physical states of being, activities, and sensations. It is broad and does not address specific diseases, disorders, or injuries, but rather health topics that impact a variety of populations. Topics include exercise, vision problems, drug use, and pain.

- **Diet**: Terms in this category are related to food and eating. Topic 13 was related to tweets mentioning fat-free, sugar-free, and milk-free food, indicating a desire for items within a certain diet. Topic 39 references specific food items as well as a discussion of meals. There is also the mention of "poisoning", indicating the bigram "food poisoning" occurred within these tweets. Topic 38 is primarily related to drinking with mentions of "water", "coffee", "alcohol", "tea", "coke", and just "caffeine." It also includes terms related to smoking as a habit.

- **Healthcare Services**: These topics are related to healthcare services as an institution. Individual topics address clinical care, hospitals, health insurance policies, and experiences related to health care. Topics 4, 6, and 10 are related to different legislation about expanded access to health care. Topics 27 and 30 are related to therapy. Topics 5, 34, 36, and 48 describe relaying and reacting to information about others' health care situations. These topics all show users' engagement within existing health care systems.

- **Mental Health**: These topics are related to a variety of mental health conditions. Topic 35 and associated tweets include terms related to body image and ability for users to appear as their identified gender. Topics 7 and 8 relate to stresses incurred by societal discrimination and lack of acceptance. Topic 32 discusses raising awareness of mental health. Topic 46 is a collection of terms related to specific diagnoses, treatments, and personal experiences. It includes "dysphoria" which refers to gender dysphoria and the sensation that a person's body does not match their identity. Topic 96 is related to the active experience of depression.

- **Symptoms**: These topics are similar to those in "general healthcare" but are more specific to a singular cause. Topics 22 and 24 are descriptions of the body aches, sore throats, and coughing related to the flu. Topic 26 is related to overexertion following exercise with cramps and leg and hip pain. Topics 28 and 54 are related to allergic reactions, and topic 16 is related to a general type of infection.

- **Transgender Healthcare**: These topics are specifically related to transgender healthcare. Topic 52 is a generalized collection of terms related to physical transitioning. Topic 25 focuses on the medical aspects of transitioning. Topics 29, 44, 23 and 41 are related to the personal and emotional experiences involved with transitioning. Topics 9, 31, and 20 are related to hormone replacement therapies as part of transitioning. Topic 1 is related to gender confirmation surgeries. Topic 12 is related to general LGBTQ health care, including issues of access and HIV.

- **Wellbeing**: These topics are related to day-to-day emotional and physical wellbeing. Topics 2 and 3 are related to recent past experiences and challenges. Topics 50 and 53 are both related to lack of sleep.

Similar to what we did in the linguistics analysis, we wanted to determine whether there was a difference in health topics discussed between TM and TW. To make this determination, we used probabilistic modeling. Specifically, for *n* tweets and *t* topics, LDA measures the probability of each topic occurring within each tweet or $P(T_k|D_j)$. We used $P(T_k|D_j)$ to find the weight of each of the topics for TM and TW separately, $WT(T_k)$. For an effective comparison, each of WTs was normalized by the sum of the weight scores of all topics:

$$\text{N\_WT}(T_k) = \frac{\sum_{j=1}^{n} P(T_k|D_j)}{\sum_{k=1}^{t} \sum_{j=1}^{n} P(T_k|D_j)}$$

If $N\_WT(T_x) > N\_WT(T_y)$, it means that users were more likely to discuss topic *x* than topic *y*. Figure 5 shows that TW users discussed general healthcare and healthcare services categories more often than TM users. Conversely, TM users discussed mental health and wellbeing more than TW users. The differences between the other categories is negligible (less than 1%) For both genders, the most discussed category is healthcare services followed by transgender healthcare. The least discussed category is diet. This finding suggests the need for more research that identifies the unique health issues of TM and TW. This analysis addresses the limitations of prior research and surveys on transgender health issues to consider TM and TW separately.

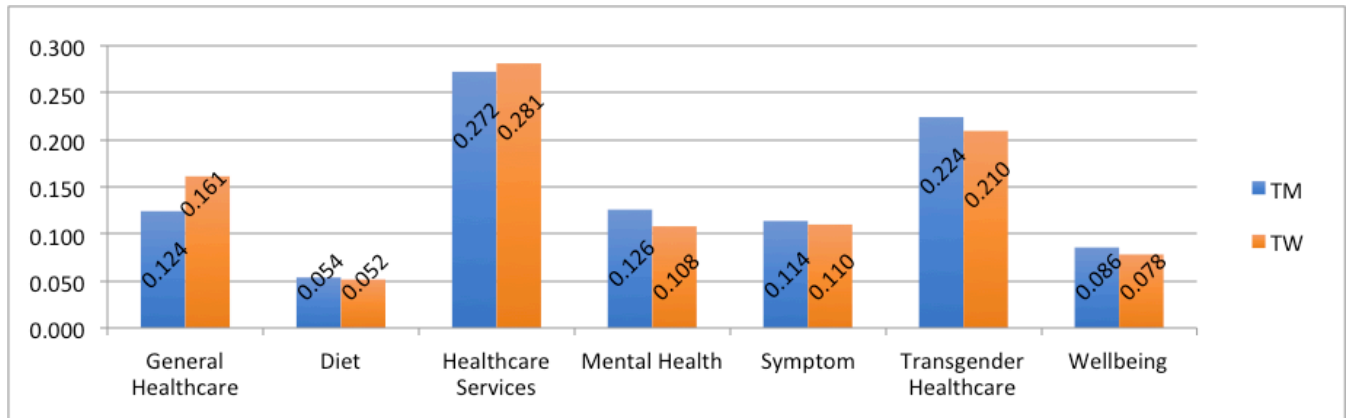

**Figure 5: Weight of Categories**

## CONCLUSION

Lack of comprehensive information on the health-related information needs of LGBTQ people represents a significant barrier to improving their health outcomes. Current studies indicate limitations with current large-scale data collection options and the need for new, macro-level, standardized approaches for collecting and analyzing data related to LGBTQ health. Transgender populations represent an especially vulnerable subset of LGBTQ individuals, with a litany of negative health outcomes and barriers to affirmative and comprehensive health services. Fortunately, social media sites like Twitter provide a great opportunity for transgender users to share their health concerns and interests.

This paper employs a computational approach to collect tweets of self-identified transgender users and explore the linguistic and semantic patterns of their tweets. Identifying the health-related concerns and interests of transgender Twitter users can help medical and public health experts better understand the needs of sexual and gender minorities.

This research has several limitations. The first is that we collected the individuals voluntarily disclosing their identities on Twitter and did not consider other transgender users. Second, although we analyzed thousands of tweets, they were shared among less than 300 transgender users. Third, all the Twitter users in this study were located in the US. Our future research plan is to increase the sample size, incorporate temporal and spatial variables, and consider other users in LGBTQ community located in both non-US countries and in the US.

**ACKNOWLEDGMENTS**

This research is supported in part by the University of South Carolina Honors College Exploration Scholars and Magellan Scholar Programs. All opinions, findings, conclusions, and recommendations in this paper are those of the authors and do not necessarily reflect the views of the funding agency.